\documentclass{emulateapj}
\usepackage{apjfonts,natbib,epsfig,color,array}
\usepackage{amsmath,tabularx} 

\usepackage{graphicx}
\newcommand{\beq}{\begin{equation}}
\newcommand{\beqa}{\begin{eqnarray}}
\newcommand{\eeq}{\end{equation}}
\newcommand{\eeqa}{\end{eqnarray}}
\newcommand{\ba}{\[\begin{aligned}}
\newcommand{\ea}{\end{aligned}\]}

\newcommand{\lsim}{\lesssim}
\newcommand{\gsim}{\gtrsim}

\newcommand{\lmk}{\left(}
\newcommand{\rmk}{\right)}

\newcommand{\lkk}{\left[} 
\newcommand{\rkk}{\right]}

\newcommand{\p}{\partial}

\shorttitle{}
\shortauthors{}

\begin{document}


\title{
Search for Galactic Civilizations Using Historical Supernovae}


\author{Naoki Seto}
\affil{Department of Physics, Kyoto University, Kyoto 606-8502, Japan}


\begin{abstract}

We study an interstellar signaling scheme which was originally proposed by Seto (2019) and efficiently links intentional transmitters to ETI searchers through a conspicuous astronomical burst, without prior communication.  Based on the geometrical and game theoretic   viewpoints, the scheme can be refined so that intentional signals can be sent and received {\it after} observing a reference burst, in contrast to the original proposal ({\it before} observing a burst). Given this inverted temporal structure,  Galactic supernovae recorded in the past 2000 years can be  regarded as  interesting guideposts for an ETI search.  While the best use period of  SN\,393 has presumably passed $\sim$100 years ago, some of the historical supernovae might allow us to  compactify the  ETI survey regions down to less than one  present of $4\pi$,  around two rings in the sky.    
\end{abstract}




\section{introduction}

An intentional signal from an extraterrestrial intelligence (ETI) has been searched for $\sim60$ years (see e.g. Drake 1961; Tarter 2001; Siemion et al. 2013), and  the Breakthrough  listen project has recently accelerated this intriguing endeavor (Lebofsky et al. 2019). However we have not achieved a definite detection yet. This might be due to   a sparseness of Galactic civilizations (see also Lingam \& Loeb 2019).  
But, we should be also aware that our current computational and observational resources might not be sufficient for exploring the multi-dimensional parameter space for SETI (Tarter et al. 2010; Wright et al. 2018). 

This difficulty at the searchers\rq{} side would be inversely deduced by the intentional senders. Therefore, in order to compactify the search parameter space, there might exist some coordinations between senders and searchers without prior communication (see e.g. Wright 2018).  If an adjustment  works efficiently, both parties can  reap benefits of significantly suppressing the required costs (e.g. transmission power, computational resources, telescopes).  In the game  theory, such a tacit adjustment in a strategy space is known as the Schelling point (Schelling 1960; Wright 2018).  

For an interstellar signaling, borrowing the phrases from Schelling (1960), uniqueness, symmetry and geometrical configuration would be relevant for converging a tacit adjustment.  In this context,  Seto (2019) pointed out that the concurrent signaling scheme could  be worth considering as   a prevailing  signaling protocol. In that scheme,  a sender transmit intentional signals by using a conspicuous astronomical burst.  Then, with the aid of the same reference  burst,  a searcher needs to examine a tiny fraction of the sky, as a function of time, irrespective of  the distances to  unknown senders.    

However, for the concurrent signaling scheme,  an intentional signal is  synchronized with the reference burst at infinite distance (Seto 2019).   At finite distances from the burst, both the senders and searchers need to complete their operations (transmissions and receptions) basically before observing the burst itself. 
Therefore, for the reference burst, we need to use  astronomical systems whose burst epochs can be accurately predicted ahead of time. A neutron star binary (NSB) with a  short orbital period ($\sim 10$~min)  could be an interesting candidate for a progenitor of a reference burst.  But  it could be difficult to  detect such a binary in the next $\sim$15 years,  before the launch of the space gravitational wave detector LISA (Kyutoku et al. 2019).  In addition, the  merger rate of Galactic NSBs might be too small,  considering the recent analysis of gravitational wave observation (Abbott et. al. 2020)

In this paper, from the perspective of the Schelling point, we reconsidered the underlying geometry of the sender-searcher-burst system.  We show that the original signaling scheme can be refined  in a way that allows us transmit and receive intentional signals after observing the reference burst.  Now, given this possibility, historical supernovae (SNe) in our Galaxy can be regarded as interesting guideposts for detecting intentional ETI signals. We pick up  six Galactic SNe recorded in the past $\sim2000$ years, and   discuss whether each of them is suitable for  a signaling reference.  We also figure out the  preferred survey directions at present, and show that the target regions could be typically less than 1\% of  the whole sky per SN.

This paper is organized as follows. In Section 2, we revisit the concurrent signaling scheme from the viewpoint of the Schelling point. 
In Section 3, we evaluate the high-priority search direction as a function of  the elapsed time after observing a reference burst.  In Section 4, we discuss the scenario of applying the historical Galactic SNe as the reference bursts. We also discuss the effects of the distance estimation errors, and make a case study for SN\,1006. Section 5 is devoted to summary and discussion.

In this paper, 
to easily make conversions between distance and time, we mainly use the light year (l.y.) for the unit of length with the conversion  1\,l.y.=0.307~pc.  We also put $c$ as the speed of light.

\section{concurrent signaling scheme}
\subsection{original proposal in Seto (2019)}

As a simple component  model, let us consider  transmissions of  short duration signals along an arbitrary straight line in the Galaxy (see Fig. 1). We introduce the coordinate $x$ for the spatial positions  on the  line. If the senders on the spatial line  (e.g., $S_1$ and $S_2$ in Fig. 1) synchronize their signals and the receiver $R_1$  knows the delivery time beforehand, the latter can collectively record the signals, irrespective of the distances to the senders. This concurrent signaling scheme would be highly advantageous for both senders and receivers.  

In principle, the timing adjustment can be done by letting the signals commonly through a specific position $x=x_D$ at a specific time $t=t_D$ (and appropriately extended in the future direction  e.g. for $S_2$ in Fig. 1).  Therefore,   the question is whether the senders and receivers can converge the datum point $(t_D,x_D)$ without prior communication. Below, we discuss this interesting possibility.

First,  we consider a rest frame covering the whole Milky-way Galaxy, ignoring  kinematical and gravitational effects. It would be reasonable to presume that the datum time  $t_D$ should be fixed by the time slice $t=t_B$ containing a conspicuous astronomical event that is well localized both temporally and spatially (a burst-like event). 

Then, the remaining question is how to choose the datum  position $x_D$  on the straight line. Here, for suppressing ambiguities, we want to keep the involved parties as small as possible (e.g. excluding an additional usage of the Galactic center).   Actually, we have the geometrically unique point on the line, given  the burst site.  That is the closest approach $C$ (with $x=x_C$) as illustrated in Fig. 2.  The combination $(t_D,x_D)=(t_B,x_C)$ is what was effectively proposed in Seto (2019). However, as understood from Fig. 2, with this choice, both the transmission  and reception of an intentional signal must be completed before observing the reference astronomical burst  (unless the burst site, sender and receiver are on the same line in this order).  More specifically, for the combination $(t_B,x_C)$, the intentional signals are synchronized with the astrophysical burst only at infinite distance (see Fig. 2).  In fact, rather than  commonalizing the datum point $(t_B,x_C)$, Seto (2019) proposed  this simple asymptotic condition   for the signal adjustment.

\begin{figure}[t]
 \begin{center}
  \includegraphics[width=48mm,angle=270,clip]{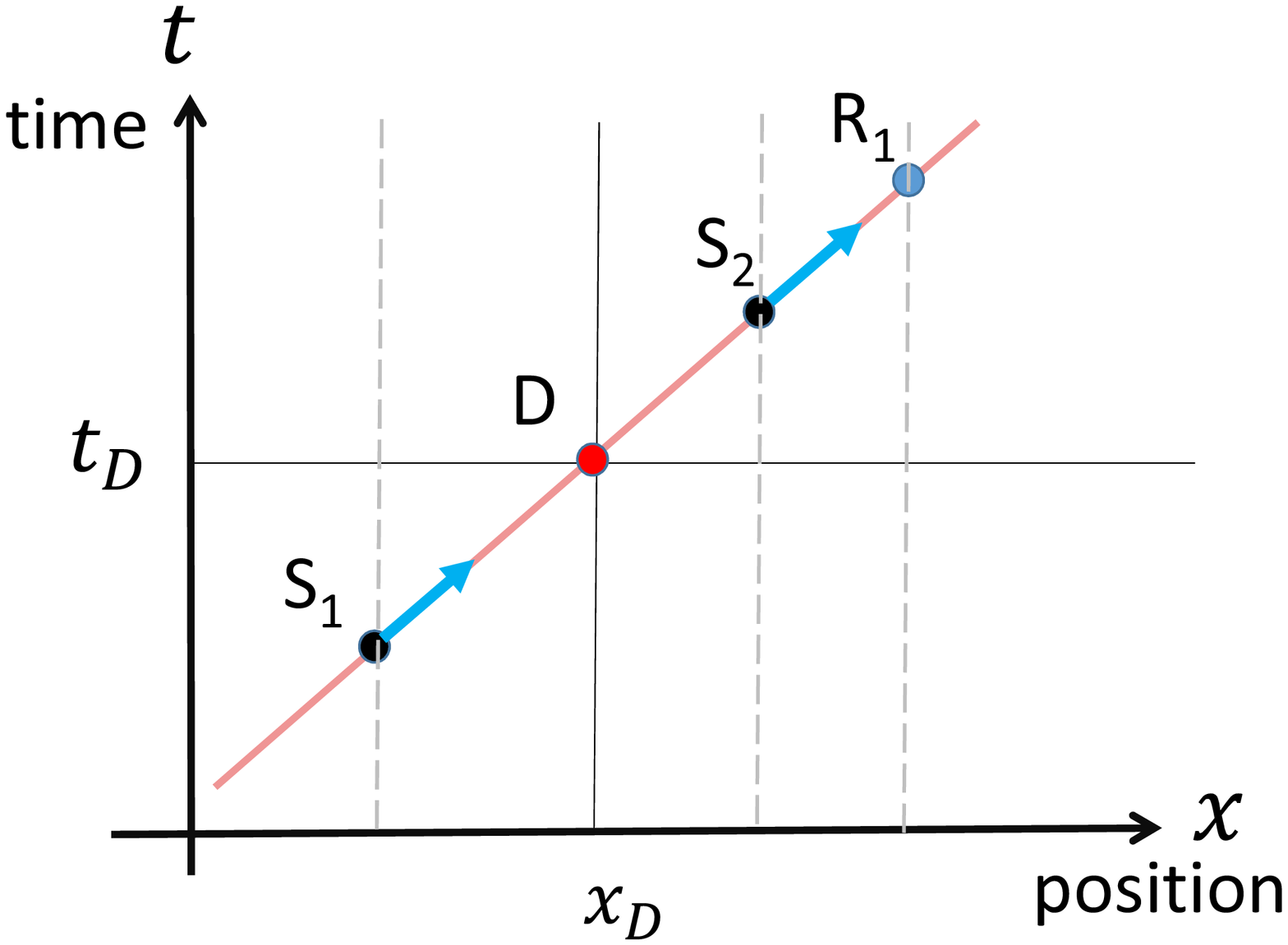}
   \caption{ 
Space-time diagram for the  concurrent signaling scheme. We consider intentional signals propagating along a straight spatial line,  corresponding to the horizontal axis with the positional coordinate $x$.  The red diagram shows a  null geodesic, and the  signal from the sender $S_2$ is emitted in phase with that from $S_1$. The receiver $R_1$ can concurrently record the signals from $S_1$ and $S_2$. This adjustment can be done by commonalizing the datum point $D$ for the senders and receivers.      }
  \label{figure:fig1}
 \end{center}
\end{figure}

Partly influenced by the broad impacts of the multi-messenger event GW170814,  Seto (2019)  pointed out that a merger of  an NSB would be ideal for the reference burst (see also Nishino and Seto 2018). 
In the signaling scheme,  to cover the whole sky,  we can use a Galactic  NSB for the time duration $2l/c\sim 10^5$ yr before the arrival of the merger signal ($l$: distance to the NSB$\sim$ size of the Galaxy).  Gravitational wave from such a short-period NSB can be easily detected by  LISA, and we can estimate the required parameters at high precision (Kyutoku et al. 2019).  But we have to wait $\sim 15$ yr before the launch of LISA,  in spite of the recent momentum of SETI related activities. 

Furthermore,  there  appeared a concern  about the appropriateness of the Galactic NSBs in relation to their merger rate.  In 2019,      Seto estimated that  the number of NSBs available for the references would be $\sim1$.   He used 
the volume averaged NSB merger rate reported by the LIGO-Virgo collaboration (Abbott et al. 2017).   However, if using  their most  recent  rate  (Abbott et al. 2020), the number of  reference NSBs in our Galaxy is  decreased by a factor of five and could be much less than unity.    Therefore, at this moment, it would be fruitful to examine applicability of astrophysical burst events other than  NSB mergers.

\subsection{refinement}

\begin{figure}[t]
 \begin{center}
  \includegraphics[width=40mm,angle=270]{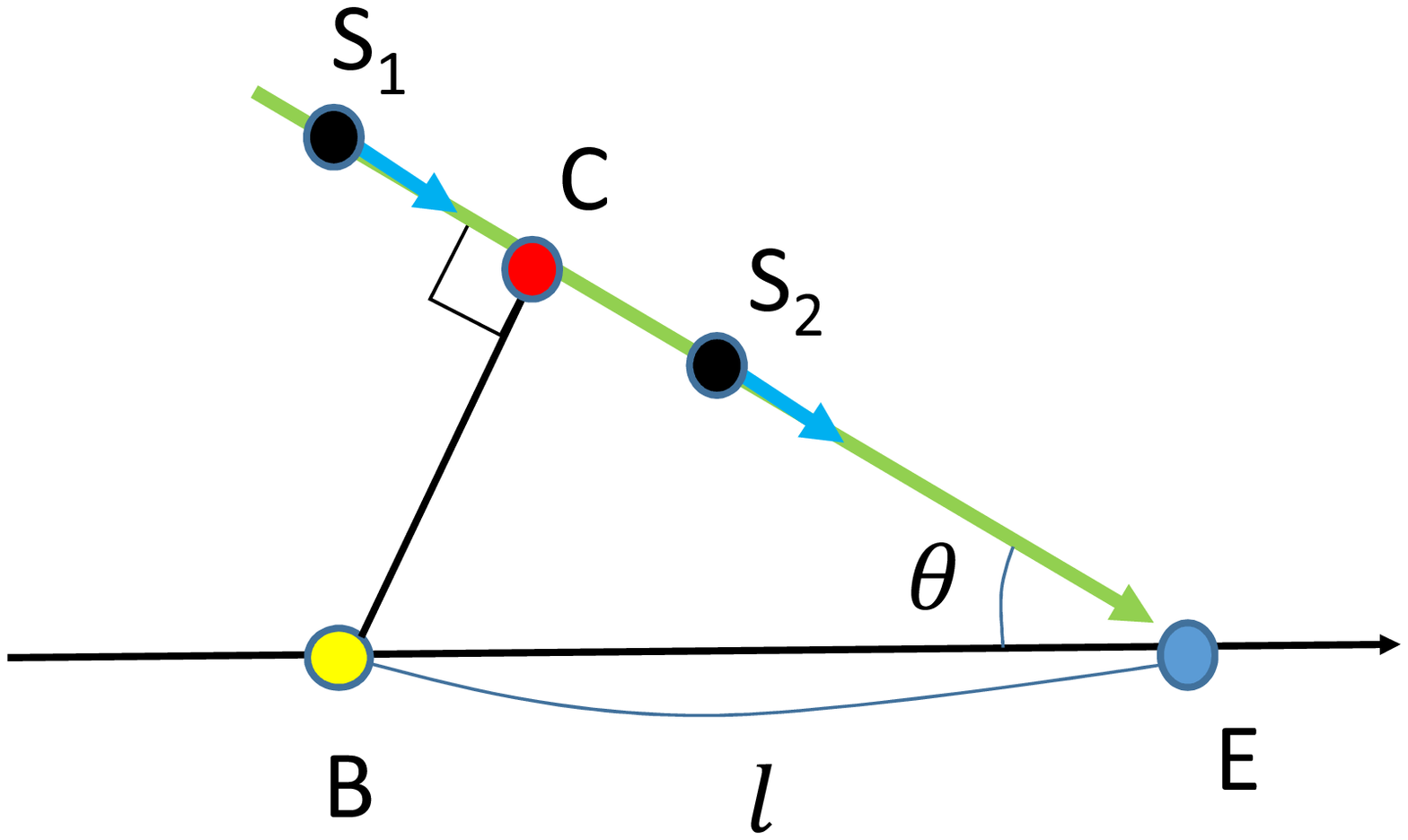}
   \caption{ Spatial configuration of the burst site (B), the  senders ($\rm S_i$) and the Earth E (as a receiver). The green line represents the spatial projection of the null geodesic mentioned in Fig. 1.   The closest approach C should be the geometrically distinctive option for the datum position  $x_D$. The closest approach distance BC would be the unique length scale existing in the system.  
}
  \label{figure:fig2}
 \end{center}
\end{figure}

Explosions of Galactic SNe are conspicuous events and  have attracted attentions of human beings at least in the past $\sim$2000 years (Stephenson \& Green 2002).   
One might wonder whether a future SN (e.g. to be observed around AD\,3000) can be currently used as the reference burst, just like a short-period NSB.
Unfortunately,  {in contrast to  an inspiral of an NSB, the physical processes up to a  SN explosion are much more complicated  
(in particular  the late-time evolution of a massive star compared with type Ia SNe)}.
 As  a result, even for advanced civilizations,    
it would not be straightforward to make a high precision prediction for the future explosion time of a SN  (e.g. with accuracy of  $\sim 1$ yr at $\sim 10^3$ yr before observing a SN).

Therefore, we want to refine the original concurrent scheme so that both the senders and receivers work only after observing a SN.  If there exists a simple solution that is also robust in the context of the Schelling point, it might be already prevailing as a tacit adjustment.  The question here can be rephrased as whether we can introduce a common time unit in order to delay the datum time $t_D$ from the original burst epoch $t_B$.

Let us take another look at Fig. 2 for  the geometry of the system composed by a signaling line and the reference burst site.  In fact, we can readily identify the geometrically  unique time interval owned by the system.  That is the closest approach distance BC divided by the speed of the light $c$.  This choice has no ambiguity and would be preferable for a tacit adjustment (Schelling 1960). In addition, the time interval  $l\sin\theta/c$ can be well estimated with  a high-precision measurement of the  burst distance $l$.

We thus expect that the closest approach $C$ would be the unique datum position $x_D$ 
 and the shifted time $t_B+q\times l\sin\theta/c$ would be a reasonable expression for the datum time $t_D$.  Here $q$ is the scaling parameter, and the choice $q=1$  would be the  primary option, in addition to the original one $q=0$.    Below, we call the choice $q=1$ by the primary mode and the  original one $q=0$ by the zero mode. 
 
 {Actually, the SN signal reaches the closest approach $C$ at the time given by  $q=1$.  Therefore, in Fig. 2, the segment CE corresponds to the portion of the signaling line where the primary mode can be transmitted after observing the reference burst. We define the corresponding length  
 \begin{eqnarray}
d_{q=1}=l \cos\theta  \label{dep}
\end{eqnarray}
 as the effective depth of the potential civilizations for the search  direction $\theta\le \pi/2$. }

Note that, for our signaling scheme,  we have the  axial symmetry with respect to the line connecting the SN site and a civilization. Therefore, at any time, the target sky directions would be rings centered at the SN or its antipodal direction.


\subsection{comments on earlier studies}

Here it might be instructive to mention some earlier discussions on the potential use of Galactic SNe for interstellar communication (Tang 1976; McLaughlin 1977; Makovetskii 1980; Lemarchand 1994, see also Hippke  2020), while they are totally different from the concurrent signaling scheme.   The basic idea of these studies is to transmit intentional signals around the observation epoch of a SN explosion. In such a method, unlike the concurrent scheme, a receiver needs to estimate the arrival times of intentional  signals individually for potential senders by using their distances.  Therefore,  except for the case when the senders are at the direction of  the SN, the  signal search would be much more demanding than the concurrent method.  Furthermore a sender needs to omni-directionally transmit their signals in a short  period of time around the observational epoch of the SN.

\section{Search and sending directions in the sky}

\subsection{search directions}

Now, we assume that some Galactic senders apply the refined  concurrent signaling scheme, employing a specific reference burst (not limited to a SN in this section).  We discuss where in the celestial sphere we should search for their intentional signals at a certain moment on the Earth.  We put $l$ as the distance between the Earth and the burst site (see Fig. 2) and define $\Delta t_E$ as the time difference (on the Earth) from the arrival epoch of the burst signal.  As explained in the previous section, we want to search for an intentional ETI signal at $\Delta t_E\ge 0$.

\begin{figure}[t]
 \begin{center}
  \includegraphics[width=78mm,angle=0]{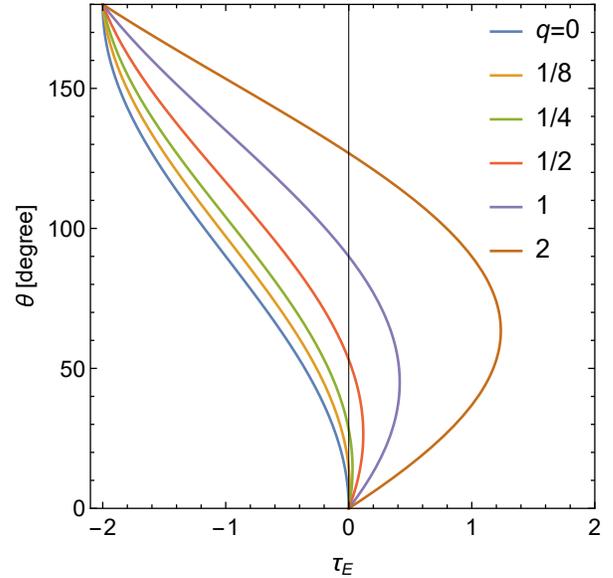}
   \caption{ The target offset angle $\theta$ away from the burst direction,  as  functions of the normalized time $\tau_E$.  The results are presented for some scaling parameters $q\ge 0$.   We must have $\tau_E\le \sqrt{1+q^2}-1$ for the existence of a solution $\theta$. }
  \label{figure:fig3}
 \end{center}
\end{figure}

Using Fig. 2, we can obtain the expression for $\Delta t_E$ as a function  of the offset angle $\theta$ 
\begin{align}
\Delta t_E&=t_B+q\times\lmk \frac{l \sin\theta}c \rmk+ \frac{l \cos\theta}c-\lmk t_B+\frac{l}c\rmk \\
&=\frac{l}c\lmk\cos\theta+q\sin\theta-1\rmk .
\end{align}
Here we introduce the normalized time $\tau_E$ by
\begin{align}
\tau_{E}&\equiv \lmk  \frac{l}c\rmk^{-1} \Delta t_E. \label{te}
\end{align}
Then we have  
\begin{align}
\tau_{E}& =\cos\theta+q\sin\theta-1 . \label{eq}
\end{align}

For a given epoch $\tau_E$ (and the parameter $q$), we can inversely  solve the offset angle $\theta$ from the burst.  More specifically, for the time range 
\begin{align}
0\le \tau_E\le \sqrt{1+q^2}-1
\end{align}
we have the two solutions $\theta=\theta_-$ and $\theta_+$ given by 
\beq
\theta_{\pm}=\arctan[q]\pm \arccos\lkk \frac{1+\tau_E}{\sqrt{1+q^2}}\rkk .  \label{kai}
\eeq
In contrast, Eq. (\ref{eq}) has no solution for $\tau_E> \sqrt{1+q^2}-1$.   The critical epoch $\sqrt{1+q^2}-1 $ is 0.41 for the primary mode $q=1$.
In Figure 3, we plot the relation (\ref{eq}) for various choices of $q$.  
For the zero mode ($q=0$) studied in Seto (2019), we have the search window $-2\le \tau_E\le 0$ (corresponding to $-2l/c\le \Delta t_E\le 0$), covering the whole sky, as mentioned in the previous section.

 For the specific angles $\theta=0$ and $180^\circ$, the closest approach  distance vanishes, and the observational epoch $\tau_E$ does not depend on the parameter $q$ in Fig. 3. Meanwhile, for $\theta=90^\circ$, the Earth becomes the closest approach $C$. 

 {Now, using  Fig. 4,  we make an intuitive explanation for the time evolution of the target search  directions  in the celestial sphere.   In this  figure,   the angle $\theta$  corresponds to  the opening angle from the burst direction, and our target directions are composed by the inner ring with $\theta=\theta_-$ and the outer ring with $\theta=\theta_+$.}  

  { At the arrival time of the burst $\tau_E=0$, we have $\theta_-=0^\circ$ and the inner ring is formed  at the direction of the burst with the effective depth $d_{q=1}=l\cos\theta_-=l$ (see Eq. (1)).  Meanwhile the outer ring emerges as the great circle at  $\theta_+=90^\circ$ away from the burst with the effective depth $d_{q=1}=l\cos\theta_+=0$. }
  
  { In the time range $0<\tau_E<\sqrt{2}-1$, the inner ring expands and the outer ring shrinks as shown in Fig. 3. Then, at the time $\tau_E=\sqrt2-1$, the two rings merge and the time window for the primary mode ($q=1$) is closed.  }

\begin{figure}[t]
 \begin{center}
  \includegraphics[width=60mm,angle=270,clip]{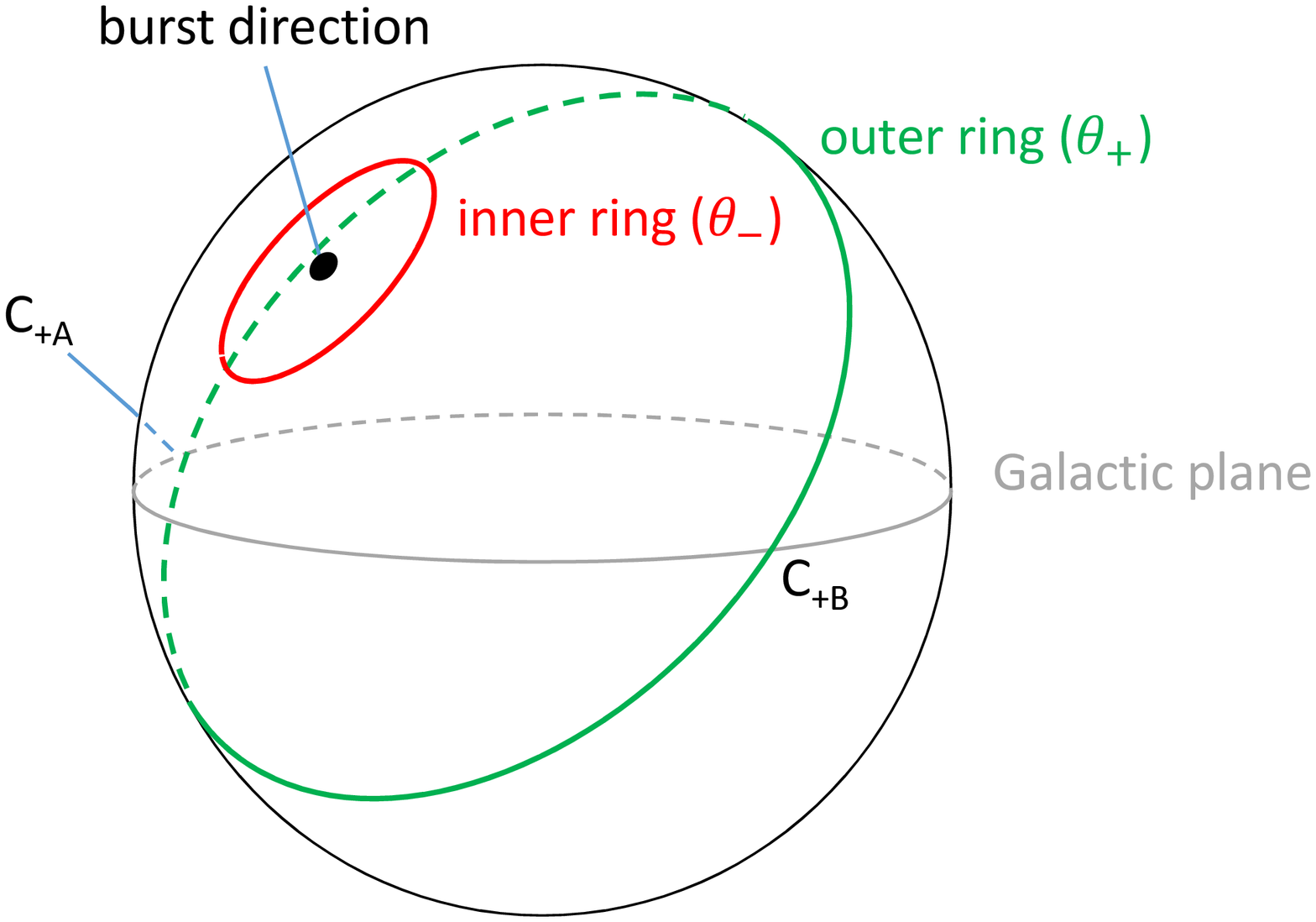}
   \caption{ The search directions in the celestial sphere   for the primary mode ($q=1$).   We have the  two separate (inner and outer)   rings centered at the burst direction with the offset angles $\theta_-$ and $\theta_+$ given by Eq. (\ref{kai}).  The inner ring expands ($\p_t\theta_->0$) with time and the outer ring shrinks ($\p_t \theta_+<0$). They  merge at  the time  $\tau_E=\sqrt2-1$ with $\theta_-=\theta_+=45^\circ$. 
    We put the intersections  between the Galactic plane in the sky and the outer ring  by $C_{+A}$ and $C_{+B}$ (see \S 4.3).  Similarly, we put $C_{-A}$ and  $C_{-B}$ for the intersections of the Galactic plane with the inner ring, but they do not exist  in this illustration.      
}
  \label{figure:fig4}
 \end{center}
\end{figure}

\subsection{sending directions}

So far, we studied the relation between $\tau_E$ and $\theta_\pm$ from the viewpoint of a receiver.  In this subsection,  we briefly discuss the directions for the signal transmission from the Earth.  In the signaling scheme, on a signaling line in Fig. 2,  the timing is identical for transmission and reception, and our expressions in the previous subsection  can be basically  used also for senders.    Correspondingly, in Fig. 4, 
the sending directions are just antipodal to the search directions, so that we can synchronize our intentional signals with the potential incoming signals.

For the primary mode ($q=1$), at the burst arrival time $\tau_E=0$, we should transmit to the antipodal direction of the burst and $90^\circ$ away from the burst (overlapping with the outer search ring only at $\tau_E=0$).  In the time interval $0\le \tau_E \le (\sqrt2-1)$, we can cover the  half-sphere centered at the antipodal direction of the burst.

At this point, we look back on the integrated structure of the refined concurrent signaling scheme. We should  notice that,  for the  propagation line in Fig. 2, the offset angle $\theta$ and the SN distance $l$  depend on the positions of the senders/receivers.  But, individually following the simple rule (\ref{eq}), we can make a globally adjusted signaling system.  Given the simplicity and effectiveness, we expect that this could be a promising Schelling point for the interstellar signaling.

\section{historical Galactic SNe}

\subsection{basic information}

In the past$\sim$2000 years, Galactic SNe have been recorded around  the world 
(Stephenson \& Green 2002). In  Table 1, we present the six historical SNe for which real-time observations were actually made and the corresponding remnants are identified relatively reliably.  Following theses two criteria,  we did not include SN\,386 and SN\,1181 in view of debates on their remnants ({see also Ritter et al. (2021) for the former with the parallax distance $\sim7500$\,l.y.}). 
We also excluded the SN resulting in  Cassiopeia A. Due to the strong  interstellar extinction, 
we do not have a secure real-time record for this SN (see also  Ashworth 1980).

 {In the second column of Table 1, we show  the Galactic coordinate for each remnant, as the combination of the Galactic longitude $l_G$ and the Galactic latitude $b_G$.  The latter will be important in \S 4.3 where we discuss the intersections of the inner/outer rings with the Galactic plane.  }

 {In the third and fourth columns of  Table 1, we presented the estimated distances $l$ to the SN sites and their relative errors $\delta l/l$ (respectively from Yuan et al. 2014; Leike et al. 2020; Winkler et al. 2003; Trimble 1973; Tian \& Leahy 2011;  Sankrit et al. 2016).  These reference values do not always represent the most conservative estimations.  
}

In the fifth column of Table 1,  for each SN, we present the normalized time $\tau_E=c\Delta  t_E/l$ at present.  SN\,393 has the largest value $\tau_E\sim0.44(>0.41)$, reflecting the long elapsed time $\Delta t_E\sim1600$~yr and the shortest distance $l$. 
Indeed, for SN\,393,  we have presumably lost accessibility to the primary mode ({even if the actual distance is $\lsim 7\%$ larger than the listed value}).   In contrast,  SN\,1604 has the smallest elapsed time $\Delta t_E\sim400$ yr and the largest distance $l$, and its normalized time is only $\tau_E=$0.025.   

In the sixth column of Table 1, except for SN\,393,  we show the two solutions $\theta_+$ and $\theta_-$ given by Eq. (\ref{kai}) for $q=1$.  The summations of the two solutions become $90^\circ$, reflecting the symmetry of Eq. (\ref{eq}) for $q=1$.  

\begin{table*}
\caption{Basic parameters for the six historical supernovae }
\centering
\begin{tabularx}{18.5cm}{XXXXXX}
\hline
              supernova                     &  Galactic coordinate        & distance $l$  [l.y.]    & uncertainty $\delta l/l$              &       $\tau_E$                   in 2021    & $\theta_+$ and $\theta_-$  for $q=1$ \\
\hline
SN\,185   & G315.4-2.3&8300 & 0.10 &0.23 &   $75.3^\circ,~ 14.7^\circ$ \\
SN\,393   &  G347.4-0.6  & 3700 & 0.01       &  0.44  & N/A   \\
SN\,1006 & G327.6+14.6 & 7100 & 0.04&  $0.14$  & $81.1^\circ,~8.9^\circ$ \\
SN\,1054 (Crab)    & G184.6-5.8 & 6500 & 0.18& 0.15  &$ 80.7^\circ,~9.3^\circ$\\
SN\,1572 (Tycho)   & G120.1+1.4 & 8970 & 0.09 &  0.051 & $ 87.1^\circ,~2.9^\circ$\\
SN\,1604  (Kepler)  & G4.5+6.8 & 16600 & 0.14  & 0.025 &  $88.5^\circ,~1.5^\circ$\\
\hline
\end{tabularx}
\vspace {2.5mm}
\end{table*}

\begin{table*}
\caption{Search regions for the primary mode ($q=1$) }
\centering
\begin{tabularx}{16.cm}{lcp{2cm}XX}
\hline
              supernova  & $ \delta \theta_\pm$ & sky fraction      & $(\alpha,\delta)$ for $C_{+A}$ and $C_{+B}$    &  $(\alpha,\delta)$ for $C_{-A}$ and $C_{-B}$          \\
\hline
SN\,185   & $1.8^\circ$ & 0.04 & $(07:44,-24.0^\circ),(18:47,-2.0^\circ)$  &   $(12:34,-62.8^\circ),(16:04,-52.5^\circ)$\\
SN\,1006 & $0.4^\circ$  & 0.008  &$ (19:20, 13.7^\circ)$,  $ (08:00, -29.7^\circ) $ & N/A\\
SN\,1054    &$1.9^\circ$ & 0.04& $(22:23,57.3^\circ),(08:54,-44.8^\circ)$ & $(05:39,31.2^\circ),(06:12,18.6^\circ)$\\
SN\,1572    & $0.28^\circ$ &0.005&  $(18:52,0.1^\circ),(06:41,5.1^\circ)$&  $(24:49,62.9^\circ),(24:05,62.4^\circ)$\\
SN\,1604   &$0.21^\circ$ &0.004&   $(21:25,50.5^\circ),(09:39,-52.5^\circ)$ & N/A\\
\hline\\
\end{tabularx}\\
The Galactic plane intersects the outer search ring at $C_{+A}$ and $C_{+B}$  and the inner search ring at  $C_{-A}$ and $C_{-B}$.  Their sky directions are presented in the equatorial coordinate $(\alpha,\delta)$.
\vspace {2.5mm}
\end{table*}

\begin{figure*}[t]
 \begin{center}
  \includegraphics[width=180mm,clip]{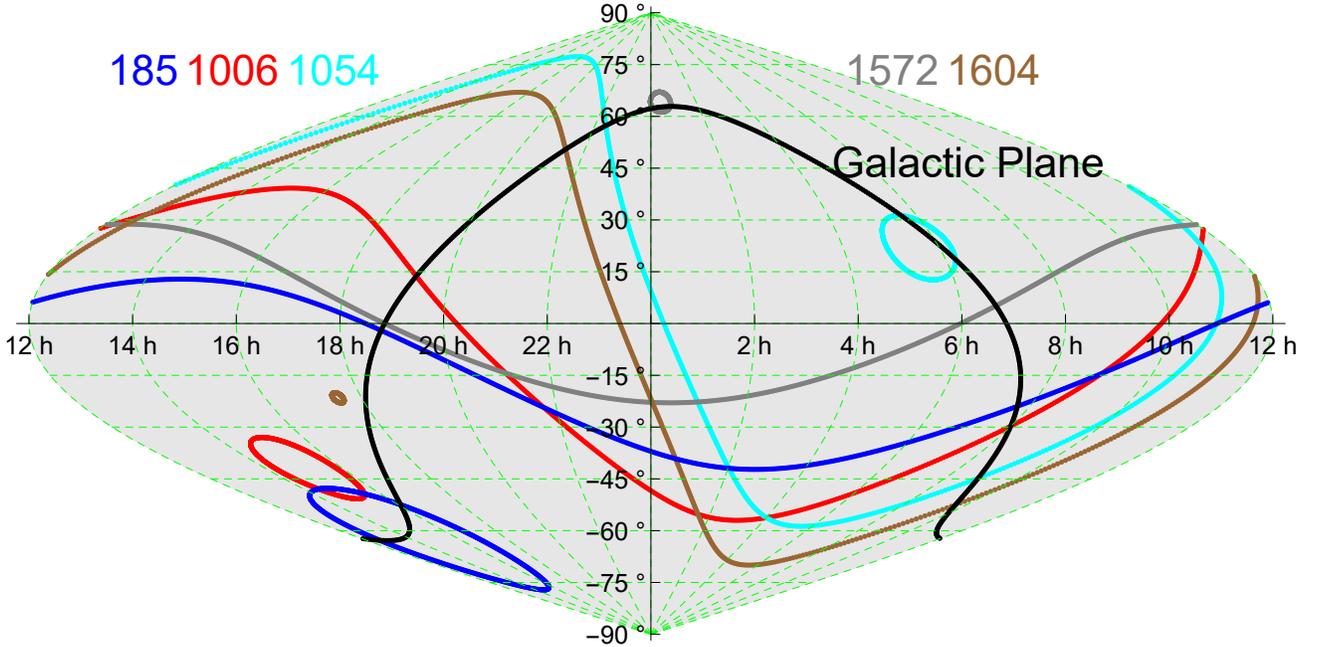}
   \caption{ The search directions associated with the five historical SNe in the equatorial coordinate.  We have the inner and outer rings for each SN.   The directions of the   SN remnants are at the centers of the inner (small) rings.  The black curve shows the Galactic plane, and its intersections with the rings are listed in Table 2.   The Galactoc center is at (17:46, $-28.9^\circ$).
}
  \label{figure:fig5}
 \end{center}
\end{figure*}

\subsection{parameter estimation errors}
 {
By using the normalized time $\tau_E=c\Delta t_E/l$ and Eq. (\ref{kai}), we can determine the offset  angles $\theta_\pm$ for the inner and  outer rings. However, because of the estimation errors $\delta l$ and $\delta (\Delta t_E)$, the angles  contain uncertainties  $\delta\theta_\pm$, and the two rings would have  the finite widths $\sim 2 \theta_\pm$. If the remnant is correctly identified for  a reference SN with an  elapsed time  $\Delta t_E>300$ yr, we  will have $\delta(\Delta t_E)/\Delta t_E < 10^{-4}$ .   Then we  can expect $\delta l/l \gg \delta(\Delta t_E)/\Delta t_E$, and the angular errors are roughly given by $\delta\theta_\pm\sim O(\delta l/l)$. 
More precisely, taking the derivative of  Eq. (\ref{eq}) for $q=1$, we obtain
\begin{eqnarray}
\delta \theta_{\pm}=\frac{\tau_E }{|\cos\theta_\pm-\sin\theta_\pm|} \frac{\delta l}l  \label{err}
\end{eqnarray}
(an identical  result for $\theta_{+}$ and $\theta_-$).
 Therefore, a high-precision distance measurement is crucial for compactifying the target sky direction.  }

In reality, some of the remnants in Table 1 contain  large relative errors $\delta l/l$.  For example, we still have $\delta l/l\sim 0.2$ for  the Crab nebula (Kaplan et al. 2008).  
In principle, the imaging parallax of the Crab pulsar could be a solid approach for its distance estimation, but the performance is limited with current radio telescopes of relatively small fields of views (see Kaplan et al. 2008 for details).   
This is due to the high system temperature caused by the hot  nebula around the pulsar and also by  the absence of reference extra-Galactic sources. In contrast, SKA is designed to have a large number of small dish telescopes (wider field views) and would  provide us with much better distance estimation to the Crab pulsar.   Here we should note that the Crab system is highly important also for a broad range of astrophysics.
The high-precision estimation of its distance is awaited not only for SETI.

RX J1713.7-3946 is considered to be the remnant of SN\,393. Leike et al. (2020) recently claimed to constrain its distance at $(3.65\pm 0.03)\times 10^3$~l.y. (corresponding to  AD\,1890-1920 for the expiration date of the primary mode search).  They used the light extinction pattern of Gaia stars around the dusty environment associated with the remnant. Using the ongoing and forthcoming astrometric projects, we might continue similar approach for SN remnants.

 {So far, we  had not taken into account the motions of involved parties. But, they cause effects such as aberration of light.  For an interstellar transmission of distance $\gsim 10^4$~l.y.,  the Galactic rotation would be the dominant contribution of $O(100{\rm km\,s^{-1}})$, and the aberration effect could be $10^{-3}$\,rad (a few arcmin).   
In future, when the distance errors are reduced down to $\delta l/l\sim 10^{-3}$, we need to consider the  aberration effect .  But the Galactic rotation can be modeled well, and the superimposed peculiar motion can be also estimated statistically (Binney \& Merrifield 1998).  Therefore, we might correct the effects related motions,  from the viewpoint of an implicit adjustment. }

\subsection{case study for SN\,1006}
 {
To provide a concrete picture of the proposed method, we make a case study for SN\,1006 (Winkler et al. 2003).  Its  estimated distance is  $l=7100$\,l.y. with an error $\delta l/l\sim 0.04$ that is relatively small in Table 1.  At present, the normalized time is $\tau_E=c(2021-1006)/l=0.14$, and we have $\theta_-=8.9^\circ$ and $\theta_+=81.1^\circ$ for the inner and outer rings of the primary mode.   The effective depths in Eq. (1) are  respectively given by $7000$\,l.y. and 1100\,l.y..  From Eq. (\ref{err}),  the widths of the rings are estimated to be $2\delta \theta_\pm \sim 0.78^\circ$.  The sky area  currently covered by the two  rings is given by
\begin{eqnarray}
\delta \Omega=4\pi \delta \theta_+ (\sin\theta_++\cos\theta_+)\sim 0.098\,{\rm sr},
\end{eqnarray}
 and its fraction to the entire sky is about $\delta\Omega/(4\pi)\sim 0.8\%$.  The accessibility to the primary mode will be lost around AD\,4000 with $1006+0.41\times 7100\sim4000$. }

On the two rings in the sky, the regions around the Galactic plane would be particularly important, given the enhanced stellar density (see also Seto 2019). The Galactic latitude of SN\,1006 is $b_G=14.6^\circ$, representing its angular separation from the Galactic plane (see Fig.  4).  Since we currently have $\theta_-<b_G<\theta_+$, only the outer ring intersects  the Galactic plane (as illustrated in Fig. 4).  In the Galactic coordinate, the two intersections $C_{+A}$ and $C_{+B}$ are given by $(l_G,b_G)=(48.4^\circ,0^\circ)$ and  $ (246.8^\circ,0^\circ)$, corresponding to  $(\alpha,\delta)= (19:20, 13.7^\circ)$ and   $ (08:00, -29.7^\circ) $ in the equatorial coordinate.  In Fig. 5,  we show the inner and outer rings with the red curves in the equatorial coordinate.  Since the preferred sending directions are antipodal to the search  directions,  it would be efficient to transmit signals to the sky around 
$(\alpha,\delta)= (07:20, -13.7^\circ)$ and   $ (20:00, 29.7^\circ) $.

We can make similar studies for other SNe listed in Table 1. For SN\,185, SN\,1054 and  SN\,1572, we have $|b_G|<\theta_-<\theta_+$ and the Galactic plane intersects also with the inner search ring (see also Fig. 5). 
In Table 2, we present  the errors $\delta\theta_\pm$ of the rings, the total sky fraction of the two rings and the equatorial coordinates for the intersections  (including those with the inner search rings for reference).    The errors $\delta\theta_\pm$ are small for SN\,1572 and SN\,1604,  due to the relation $\delta\theta_\pm\propto \tau_E$ (see Eq. (\ref{err})).

\section{summary and discussion}

A tacit adjustment on intentional signaling can significantly reduce the required costs both  for the senders and searchers and could be actually prevailing in our Galaxy. In this context,  Seto (2019) pointed out the possibility of the concurrent signaling scheme that is  controlled by astronomical bursts such as BNS mergers. But, in the original scheme, we need to complete operations basically before observing the reference bursts themselves.  This would severely limit astrophysical systems suitable for reference bursts.

In this paper, we reconsider the underlying geometry of the system and identify the unique time unit corresponding to the closest approach distance between the reference burst and the signal line. We can slide the datum time $t_D$ (in Fig. 1) with a parameter $q$.  Its primary option  would be $q=1$ from the viewpoint of  the Schelling point.    
Under this scenario,  we can search (or transmit) intentional signals after observing a reference burst. 
More specifically, we can determine the target directions by using the distance to the reference $l$ and the elapsed time $\Delta t_E$ after observing  the burst.

Now, the historical records of Galactic SNe in the past $\sim 2000$ years can be regarded as invaluable assets for SETI with the accurate knowledge on the elapsed times  $\Delta t_E$. 
We found that for SN\,393, the accessibility to the primary mode ($q=1$) was likely to be lost about 100 years ago, but we still have  multiple SNe applicable for the primary mode search.

Given the estimated  Galactic SN rate ($\sim1$ per 100 years; Tammann et al. 1994), a SN might be  suitable for an intermediate-distance transmission  (e.g. $l\lsim 10^4$~l.y.) in our Galaxy. Otherwise we have too many references at any moment, and cannot compactify the target sky directions. Considering the time duration $\sim l/c$ to cover a large fraction of the sky, the selection of a reference burst might depend also on the self-estimated lifetime of the transmitting civilization. 

Here we should notice that astronomical  {events} other than SNe  might be also used for the signaling reference, particularly if the two parameters $l$ and $\Delta t_E$ can be estimated accurately {(e.g. those associated with orbital motions around Sgr A*)}. Furthermore, we might not need to exclude the possibility of using a negative $q$ for predictable bursts such as BNS mergers.  

In this paper, we have concentrated on the geometrical aspects of the system. This is partly because of the less ambiguous nature of the consideration and the resultant compatibility with the Schelling point argument. But it might be fruitful to discuss possibilities of other tacit adjustments from different viewpoints.

\begin{acknowledgments}
 The author would like to thank the anonymous reviewer for his/her comments and suggestions. This work is supported
by JSPS Kakenhi Grant-in-Aid for Scientific Research
(Nos. 15K65075, 17H06358).  
\end{acknowledgments}

\end{document}